\documentclass[aps,prl,twocolumn,groupedaddress,showpacs]{revtex4}
\usepackage{graphicx}
\newcommand{\shortv}[1]{}
\newcommand{\longv}[1]{#1}
\begin{document}
\title{Open circular billiards and the Riemann hypothesis}
\author{L. A. Bunimovich}
\affiliation{Southeast Applied Analysis Center, Georgia Institute of Technology, Atlanta GA 30332-1060, USA}
\author{C. P. Dettmann}
\affiliation{School of Mathematics, University of Bristol, Bristol BS8 1TW, UK}
\date{\today}
\begin{abstract}
A comparison of escape rates from one and from two holes in an experimental
container (e.g. a laser trap) can be used to obtain information about
the dynamics inside the container. If this dynamics is simple enough one can 
hope to obtain exact formulas.  Here we obtain exact formulas for escape
from a circular billiard with one and with
two holes.  The corresponding quantities are expressed as sums over zeroes
of the Riemann zeta function.
Thus we demonstrate a direct connection between recent experiments and
a major unsolved problem in mathematics, the Riemann hypothesis.
\end{abstract}
\maketitle
Billiard systems, in which a point particle moves freely except for
specular reflections
from rigid walls, permit close connections between rigorous mathematics
and experimental physics.  Very general physical situations, in which
particles or waves are confined to cavities or other homogeneous regions, are
related to well understood billiard dynamical systems, directly for particles
and via semiclassical (short wavelength) theories for waves.  Precise
billiard experiments have used microwaves in metal~\cite{S91,SS92} and
superconducting~\cite{Gea92} cavities and with wave guides~\cite{PRSB00},
visible light reflected from mirrors~\cite{SZOL01}, phonons in
quartz blocks~\cite{EGLNO96}, electrons in
semiconductors confined by electric potentials~\cite{Sea91,B99}, and atoms
interacting with laser beams~\cite{MHCR01,FKCD01,AKGD}.
Experiments employ both two and three~\cite{Dea02}
dimensional geometries, and both closed and open systems.  For instance,
escape of cold atoms from a laser trap with a hole was studied
in~\cite{MHCR01,FKCD01}.  Closed systems
exhibit energy level distributions and scarring of wavefunctions as
predicted by semiclassical~\cite{M99} and random matrix~\cite{FSV03} theories.
Open systems exhibit unexpected and incompletely understood phenomena such as
fractal conductance fluctuations~\cite{B99,Tea97,T02,CSGFBR}.
For both closed and open systems, the behavior depends crucially on the
classical dynamics, which can be tuned to be integrable, chaotic, or
mixed~\cite{MHCR01,FKCD01,Tea97,B01}.  Escape rate is a characteristic of
open billiards which is both experimentally accessible~\cite{MHCR01,FKCD01}
and important for transport properties of many systems.  In this
\shortv{Letter}\longv{paper} we propose to ask what may be understood about
the intrinsic dynamics of billiards using only this experimental
escape information, specifically by comparing systems with one and two holes.
The two hole escape rate is not measured in the experimental literature, but
presents no fundamental difficulties.  We naturally begin with the simplest
of shapes, the circle, and find remarkable exact expressions based on the
most famous unsolved mathematical problem, the Riemann Hypothesis. 

At long times, the probability of a particle remaining in an integrable
billiard with a hole is well known to exhibit power law decay, in contrast to
exponential decay from strongly chaotic billiards~\cite{BB},
however the coefficient of the power (``escape rate'') in the integrable case
has not been computed
exactly to our knowledge.  \longv{Numerical simulations can lead not only to
uncertainty in the coefficient, but also misleading conclusions, for example
the power law decay of an integrable billiard can be masked by an exponential
term at short times.}\shortv{Numerical simulations can be misleading, for
example a power law decay at long times can be masked by an exponential
term at short times.}
  
Here we consider the circle billiard, which is integrable due to angular
momentum conservation.  Some three dimensional cases, namely the cylinder
and sphere, can be treated analogously.  The circle billiard of unit radius
has collisions defined by the angle around the circumference
$\beta\in(-\pi,\pi]$ and the angle
between the outward trajectory and the normal $\psi\in(-\pi/2,\pi/2)$.
The billiard map is then $(\beta,\psi)\to(\beta+\pi-2\psi,\psi)$
where angles are taken modulo $2\pi$ as usual.  The time between collisions
is $T=2\cos\psi$.

The dynamical evolution is of two types depending on the value of $\psi$.
For $\psi=\psi_{m,n}\equiv\pi/2-m\pi/n$ with $m<n$ coprime integers, the
trajectory \shortv{has} \longv{is periodic with}
period $n$ and the $\beta$ values are equally spaced at
intervals of $2\pi/n$.  For $\psi$ which are irrational multiples
of $\pi$, the trajectory is uniformly distributed in $\beta$
\longv{ (see for example Prop. 4.2.1 of Ref.~\cite{KH})}.

\begin{figure}
\begin{picture}(200,180)(80,270)
\includegraphics[width=380pt]{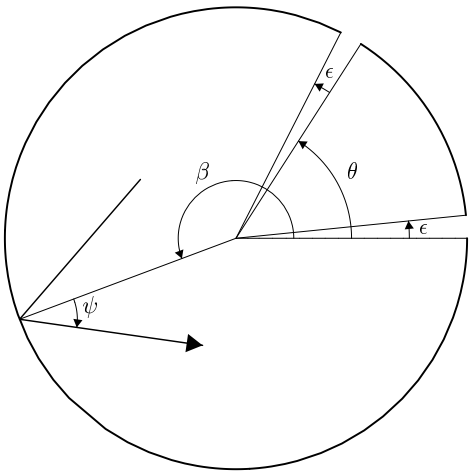}
\end{picture}
\caption{Geometry of the billiard.\label{f:circ}}
\end{figure}

For the escape problem the billiard is filled with a
uniform density of particles normalised to unity given by
$\cos\psi d\beta d\psi/(4\pi)$ at the \longv{boundary; this is the usual
equilibrium measure for two dimensional billiards.  Note that for the circle,
any function of $\psi$ leads to an invariant measure since $\psi$ is
a constant of motion.}\shortv{boundary.}
Two (possibly overlapping) holes are placed at the boundary at
$\beta\in[0,\epsilon]$ and
$\beta\in[\theta,\theta+\epsilon]$; the one hole problem is simply
$\theta=0$.  The number of collisions to escape is some function
$N(\beta_0,\psi_0)$ (possibly infinite)
of the initial conditions and the time to escape is
$t=NT=2\cos\psi N(\beta_0,\psi_0)$.
\longv{Considering the continuous and discrete
time problems leads to escape times differing at most by the maximal free
path of 2; only the discrete time calculation will be required for the
leading (in time) behavior of escape, that is, the coefficient of $1/t$ for
the density remaining at time $t$.
The geometry of the billiard is shown
in Fig.~\ref{f:circ}.}
\shortv{See Fig.~\ref{f:circ}.}

\longv{
From above, the only trajectories which do not escape are some of the
periodic orbits with
period $n<2\pi/\epsilon$.  For an arbitrary aperiodic (finite time) orbit
we define the period $n$ to be the smallest positive integer for which
$|\phi_n-\phi_0|\leq\epsilon$.  For the largest possible value of
$n<2\pi/\epsilon$, the $n+1$ points $\phi_0, \phi_1, \ldots \phi_{n}$
automatically have a minimum distance $\leq\epsilon$ on the circle,
so the inequality is satisfied by this definition.  In units of $n$
collisions, the particle's orbit precesses slowly enough to be captured
by one of the holes, so at very large values of $t$, only slightly precessing
orbits remain.
}

Now we compute the density remaining near periodic orbits $(\beta_0,m,n)$
at long time $t$.  Such a long lived trajectory has $\psi=\psi_{m,n}+\eta$
for $\eta\ll\epsilon$.  A prime will indicate values taken
modulo $2\pi/n$ and lying in $[0,2\pi/n)$.
Thus the dynamics is now $\beta'\to\beta'-2\eta$
and escape takes place when this passes one of the values $\epsilon$ or
$\epsilon+\theta'$ for $\eta>0$, or $0$ or $\theta'$ for $\eta<0$.
The set of initial conditions for which escape takes at least time $t$,
hence $t/(2\cos\psi_{m,n})$ collisions is thus
\begin{equation}
\beta'_0\in\left(\epsilon+\frac{\eta t}{\cos\psi_{m,n}},\theta'\right)
\bigcup
\left(\theta'+\epsilon+\frac{\eta t}{\cos\psi_{m,n}},\frac{2\pi}{n}\right)
\end{equation}
for $\eta>0$ and similar expressions for $\eta<0$.  Integrating over
this region, including the $\cos\psi$ weighting for the invariant measure
and a factor $n$ to account for the equivalence modulo $2\pi/n$, and summing
over all non-escaping periodic orbits gives
\begin{eqnarray}\nonumber
P(t)&\sim&\frac{1}{4\pi}\sum_{m,n}
\frac{n\left[g\left(\frac{2\pi}{n}-\theta'-\epsilon\right)
+g\left(\theta'-\epsilon\right)\right]}{t}\sin^2\frac{\pi m}{n}\\
g(x)&=&\left\{\begin{array}{cc}x^2&x>0\\0&x\leq 0\end{array}\right.
\end{eqnarray} 
where the sum is restricted to $0\leq m<n<2\pi/\epsilon$ for coprime $m,n$,
and faster decaying terms in $t$ have been neglected \longv{(for example in
approximating $\cos\psi$ by a constant near each periodic orbit)}.
Note that $n=1$ is included, but gives no contribution.
Now we write the $\sin$ functions as sums of exponentials and
apply the Ramanujan identity~\cite{D}
\begin{equation}
\sum_{\shortstack{m=0\\gcd(m,n)=1}}^{n-1} e^{2\pi i m/n}=\mu(n)
\end{equation}
where $\mu$ is the M\"obius function, defined by $\mu(1)=1$, $\mu(p)=-1$
for primes $p$, $\mu(mn)=\mu(m)\mu(n)$ if $gcd(m,n)=1$, otherwise $\mu(mn)=0$.
The result is
\begin{eqnarray}\nonumber
P_\infty&\equiv&\lim_{t\rightarrow\infty}
tP(t)=\frac{1}{8\pi}\sum_{n=1}^\infty n(\phi(n)-\mu(n))
\\&&\times\left[g\left(\frac{2\pi}{n}-\theta'-\epsilon\right)
+g\left(\theta'-\epsilon\right)\right]\label{e:tP(t)}
\end{eqnarray} 
Here $\phi(n)$ is the Euler totient function, giving the number of
positive integers $m\leq n$ with $gcd(m,n)=1$.  Conventionally $\phi(1)=1$.
Of course $P_\infty$ depends on both $\epsilon$ and $\theta$, but this
is suppressed for brevity.

$P_\infty$ is a finite sum, piecewise smooth when considered as a
function of $\epsilon$ and/or $\theta$ with the number of terms unbounded
in the limit of small $\epsilon$.  For small
holes we can extract the asymptotic behavior using Mellin transforms.
Write $\tilde{P}(s)=\int_0^{\infty}P_\infty\epsilon^{s-1}d\epsilon$ 
and then
$P_\infty=\frac{1}{2\pi i}\int_{c-i\infty}^{c+i\infty}\epsilon^{-s}
\tilde{P}(s)ds$ 
gives the required asymptotic series as a sum of residues.
\longv{The constant $c$
is determined by convergence of the first integral.  This integral is
finite ($P_\infty=0$ when $\epsilon>\pi$) and so converges for sufficiently
large $s$, leading to a constant $c$ to the right of all poles of
$\tilde{P}(s)$.  For sufficiently large $s$ convergence of the sum is
uniform in $\epsilon$, so interchanging}
\shortv{Interchanging} the sum and the first integral,
integrating over $\epsilon$
and substituting $\theta'=f(n\theta/(2\pi))2\pi/n$ where $f$ indicates
the fractional part, we find
\begin{eqnarray}\nonumber
P_\infty&=&\frac{1}{2\pi i}\int_{c-i\infty}^{c+i\infty}
\frac{ds\epsilon^{-s}(2\pi)^{s+1}}{2s(s+1)(s+2)}
\sum_{n=1}^{\infty}\frac{\phi(n)-\mu(n)}{n^{s+1}}
\\&&\times\left\{\left[1-f\left(\frac{n\theta}{2\pi}\right)\right]^{s+2}
+f\left(\frac{n\theta}{2\pi}\right)^{s+2}\right\}\label{e:contour}
\end{eqnarray}

Now we consider the case of rational angles $\theta=2\pi r/q$ where
$gcd(r,q)=1$; the $r=0$, $q=1$ case gives a single hole.  In this case the
sum over $n$ splits into conjugacy classes modulo $q$ and the fractional
parts are known rational numbers which depend only on the conjugacy class.

\longv{
Let us consider the required sum
\begin{equation}
\sum_{n\equiv a \pmod{q}}\frac{\phi(n)-\mu(n)}{n^{s+1}}
\end{equation}
by first dividing all quantities through by
$b=gcd(a,q)$, writing $n'=n/b$, $a'=a/b$ and $q'=q/b$.  We then have
\begin{equation}
\sum_{n'\equiv a' \pmod{q'}}\frac{\phi(bn')-\mu(bn')}{(bn')^{s+1}}
\end{equation}
with $a'$ and $q'$ coprime.}
\shortv{
To evaluate the sum over $n$ we first divide all quantities through by
$b=gcd(a,q)$, writing $n'=n/b$, $a'=a/b$ and $q'=q/b$.
}
Now we consider Dirichlet characters~\cite{D} $\chi(n')$, defined as follows.
The conjugacy classes modulo $q'$ which are coprime to $q'$ form an
abelian group under multiplication, of order $\phi(q')$.  Since the
group is abelian and finite, there are $\phi(q')$ irreducible representations
$\chi$ in which each $n'$ coprime to $q'$ is represented by a complex
root of unity $\chi(n')$ satisfying $\chi(m')\chi(n')=\chi(m'n')$;
$\chi(n')=0$ if $n'$ and $q'$ have a common factor.

\longv{
A central result in representation theory is the orthogonality theorem,
here~\cite{D}
\begin{equation}
\frac{1}{\phi(q')}\sum_\chi \bar\chi(a')\chi(n')=\delta_{a',n'}
\end{equation}
where $\delta_{a',n'}=1$ if $a'\equiv n' \pmod{q'}$, zero otherwise, and
the bar indicates complex conjugation.
}

\begin{figure}
\begin{picture}(200,190)(10,5)
\includegraphics[width=200pt]{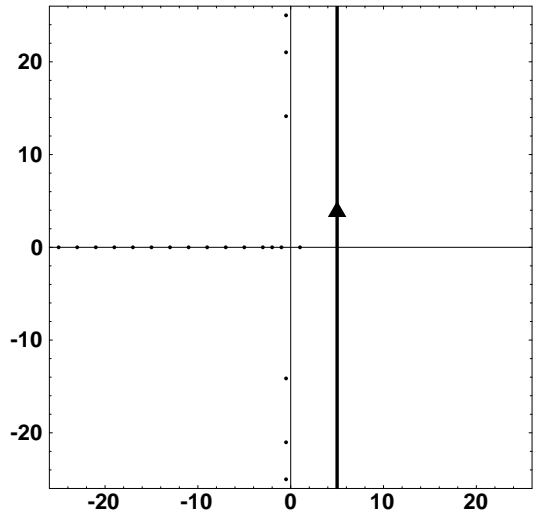}
\end{picture}
\caption{Poles of $\tilde{P}(s)$ (Eq.~\protect\ref{e:Ps})
for $q=1$ (one hole) at -2, odd
integers less than or equal to 1, and nontrivial values with real part
$-1/2$ assuming the Riemann hypothesis.  The contour of
Eq.~\protect\ref{e:contour} is given.
\label{f:poles}}
\end{figure}

Inserting the orthogonality relation \shortv{for characters~\cite{D}}
into the sum allows $n'$ to be summed over all integers
and decomposed into prime factors $n'=\prod_pp^{\alpha_p}$,
\begin{eqnarray}\nonumber
&&\sum_{n\equiv a \pmod{q}}\frac{\phi(n)-\mu(n)}{n^{s+1}}
\shortv{
=\!\!\!\!\sum_{n'\equiv a' \pmod{q'}}\!\!\!\!
\frac{\phi(bn')-\mu(bn')}{(bn')^{s+1}}
}
\\
&&=\frac{1}{\phi(q')}\sum_\chi\bar\chi(a')\sum_{n'=1}^\infty
\chi(n')\frac{\phi(bn')-\mu(bn')}{(bn')^{s+1}}\\\nonumber
&&\chi(n')=\prod_p\chi(p)^{\alpha_p},\quad 
\phi(bn')=\phi(b)\prod_{p|n',\;\; p\!\not\,|b}(1-p^{-1})\\
&&\mu(bn')=\left\{\begin{array}{cc}\mu(b)\prod_{p}(-1)^{\alpha_p}&
c^2\not|bn'\quad\mbox{for all $c>1$}\\0&\mbox{otherwise}\end{array} \right.
\end{eqnarray}
\shortv{Here, the bar indicates complex conjugation.}
The condition on $\mu$ is taken into account by setting $\alpha_p=0$ if
$p|b$ and summing $\alpha_p=0,1$ otherwise.  The sum for $\phi$ is given
by the $\alpha=0$ term together with a geometric series.  The result is
\begin{eqnarray}
&&\sum_{n\equiv a \pmod{q}}\frac{\phi(n)-\mu(n)}{n^{s+1}}\\\nonumber&&=
\frac{1}{b^{s+1}\phi(q')}\sum_\chi\frac{\bar\chi(a')(\phi(b)L(s,\chi)-\mu(b))}
{L(s+1,\chi)\prod_{p|b}(1-\chi(p)p^{-s-1})}
\end{eqnarray}
where characters are taken modulo $q'$ and 
\begin{equation}L(s,\chi)=\sum_{n=1}^{\infty}\frac{\chi(n)}{n^s}
=\prod_p(1-\frac{\chi(p)}{p^s})
\end{equation}
is the Dirichlet $L$ function, which in the case $q'=1$ (ie $\chi(n)=1$ for
all $n$) reduces to the
Riemann zeta function. \longv{For each $q'$ there is a trivial character for
which $\chi(a')=1$ for all $a'$ coprime to $q'$; this gives the same
product as the Riemann zeta function, excluding primes dividing $q'$,
thus
\begin{equation}
L(s,1)=\zeta(s)\prod_{p|q'}(1-p^{-s})
\end{equation}
}
Our first main result is the exact expression for the probability $P(t)$
of remaining in the unit circular
billiard with two holes $[0,\epsilon]$ and $[2\pi r/q,2\pi r/q+\epsilon]$ is
thus
\begin{eqnarray}\nonumber
&&\lim_{t\rightarrow\infty} tP(t)=\sum_j{\rm res}_{s=s_j}\tilde{P}(s)
\epsilon^{-s}\\\nonumber 
\tilde{P}(s)&=&\frac{(2\pi)^{s+1}}{2s(s+1)(s+2)}
\sum_{a=1}^q\frac{(1-f(\frac{ap}{q}))^{s+2}+f(\frac{ap}{q})^{s+2}}
{b^{s+1}\phi(q')}\\
&&\times\sum_{\chi}\frac{\bar\chi(a')(\phi(b)L(s,\chi)-\mu(b))}
{L(s+1,\chi)\prod_{p|b}(1-\chi(p)p^{-s-1})}\label{e:Ps}
\end{eqnarray} 
where, as above, $b=gcd(a,q)$, $a'=a/b$, $q'=q/b$ and the characters are taken
modulo $q'$.
In performing the contour integral~(Fig.~\ref{f:poles})
we recall that the contour
lies to the right of all poles of the integrand ($c>1$) and that a semicircular
arc to the left may be added that avoids the poles and for which the
integral vanishes in the limit of infinite radius.
\longv{For large negative real part the integrand
is controlled by the $(2\pi)^{s+1}$ and for large imaginary part the
integrand is controlled by the cubic prefactor in the denominator.}

We now consider specific values of $q$; recall that the angle 
separating the holes is $\theta=2\pi r/q$.  Thus $q=1$ is a
single hole and $q=2$ is two opposite holes.  Experimentalists
could especially notice a contrast between small values of $q$ in which the
angles are simple rational multiples of $\pi$ and the escape rate is
expressed in terms of only a few $L$-functions (in fact only the Riemann
zeta function for $q=1,2,3,4,6$), and angles which are not close to
rational numbers with small denominators.  \longv{Particular values of
$\tilde{P}(s)$ are given in table~\ref{t:P(s)} and residues of the trivial
poles at integer $s$ are given in table~\ref{t:poles}.}
\shortv{Particular values of $\tilde{P}(s)$ and its residues are given in
table~\ref{t:P(s)}.}
\begin{table}
\begin{tabular}{|c|c|\shortv{ccc|}}\hline
$q$&$\tilde{P}(s)(q/2\pi)^{s+1}$\shortv{&$1$&$-1$&$-2$}\\\hline
1&$\frac{\zeta(s)-1}{2s(s+1)(s+2)\zeta(s+1)}$
\shortv{&2&$-\frac{13}{12}$&$\frac{3}{2\pi}$}\\
2&$\frac{\zeta(s)}{s(s+1)(s+2)\zeta(s+1)}$
\shortv{&1&$-\frac{1}{6}$&0}\\
3&$\frac{3^s(7\zeta(s)+2^{s+2}(\zeta(s)-1)+2)
-\zeta(s)(2^{s+2}+1)}{2s(s+1)(s+2)(3^{s+1}-1)\zeta(s+1)}$
\shortv{&1&$-\frac{1}{4}-\frac{5\ln2}{9\ln3}$&$\frac{3}{4\pi}$}\\
4&$\frac{2^s(13\zeta(s)+3^{s+2}(\zeta(s)-1)+3)
-\zeta(s)(3^{s+2}+5)}{4s(s+1)(s+2)(2^{s+1}-1)\zeta(s+1)}$
\shortv{&1&$-\frac{1}{3}-\frac{11\ln3}{16\ln 2}$&$\frac{3}{\pi}$}
\longv{\\
&\raisebox{-8pt}[-8pt][8pt]{$\scriptstyle[(\pi/3)^{s+1}(6^s+8.12^s-25.30^s+
(1-3.2^s-13.3^s-8.4^s$}\\6&$\scriptstyle
+25.5^s+27.6^s-25.10^s+8.12^s-25.15^s+25.30^s)\zeta(s))
]$\\&\raisebox{4pt}[4pt][-4pt]{$\scriptstyle\times[
2s(s+1)(s+2)(2^{s+1}-1)(3^{s+1}-1)\zeta(s+1)]^{-1}
$}}
\\\hline
\end{tabular}
\caption{The function $\tilde{P}(s)(q/2\pi)^{s+1}$ (see Eq.~\protect\ref{e:Ps})
\shortv{and some of the residues of $\tilde{P}(s)$}
for $q=1,2,3,4\longv{,6}$ and $r=1$.
\label{t:P(s)}}
\end{table}
\longv{
\begin{table}
\begin{tabular}{|c|cccc|}\hline
$q$&\multicolumn{4}{c|}{$s$}\\
&1&-1&-2&-3\\\hline
1&2&$-\frac{13}{12}$&$\frac{3}{2\pi}$&$\frac{119}{5760\pi^2\zeta'(-2)}$\\
2&1&$-\frac{1}{6}$&0&$-\frac{1}{720\pi^2\zeta'(-2)}$\\
3&1&$-\frac{1}{4}-\frac{5\ln2}{9\ln3}$&$\frac{3}{4\pi}$&
$\frac{49}{5120\pi^2\zeta'(-2)}$\\
4&1&$-\frac{1}{3}-\frac{11\ln3}{16\ln 2}$&$\frac{3}{\pi}$&
$\frac{109}{1620\pi^2\zeta'(-2)}$
\longv{\\
&&\raisebox{-8pt}[-8pt][8pt]{$
\scriptstyle[5\ln5(10\ln3-7\ln5)+\ln2(55\ln5-76\ln3)$}&&\\
6&1&$\scriptstyle+(10\ln5-8\ln2)
(7\ln\epsilon+12\zeta'(-1))]$&$-\frac{3}{2\pi}$&
$-\frac{79}{6400\pi^2\zeta'(-2)}$\\
&&\raisebox{4pt}[4pt][-4pt]{$\scriptstyle\times[72\ln2\ln3]^{-1}$}&&}
\\\hline
\end{tabular}
\caption{Some residues of $\tilde{P}(s)\epsilon^{-s}$
given in table~\protect\ref{t:P(s)} divided by the factor $\epsilon^{-s}$.
\longv{The $\ln\epsilon$ appears for $q=6$ due to a double pole at $s=-1$.}
There are also poles for further negative odd $s$, and along the
critical line ${\cal R}e\;s=-1/2$.
\label{t:poles}.}
\end{table}
}
Some points to note are that the leading behavior at $s=1$, that is, of
order $\epsilon^{-1}$ is purely given by the total size of the holes.  The
$q=1$ case is twice as large since there is only a single hole in this case.
\longv{
The next possible term would be $s=0$, however the residue is zero since
the $\zeta(s+1)$ in the denominator has a pole there.  The $s=-1$ term is
more complicated for $q=6$, involving a term in $\epsilon\ln\epsilon$ due
to the double pole in this case.  The $s=-2$ terms are remarkably simple,
and are due to the $s+2$ in the denominator.
}
For these values of $q$ the second to leading order
terms, which are not given in the table, come from the nontrivial zeroes of the
$\zeta(s+1)$, and for multiplicity $m$ are of order
$\sqrt{\epsilon}(\ln\epsilon)^{m-1}$
if all zeroes of $\zeta(s)$ lie in ${\cal R}e\;s\leq1/2$.  The latter is
a statement of the Riemann hypothesis~\cite{RH}.  An alternative formulation
stating that all the nontrivial zeroes of $\zeta(s)$ have ${\cal R}e\;s=1/2$ is
easily shown to be equivalent using the functional equation~\cite{RH}
relating $\zeta(s)$
and $\zeta(1-s)$.  Thus this celebrated unsolved problem is equivalent to
either of
\begin{eqnarray}\lim_{\epsilon\to 0}\lim_{t\to\infty}
\epsilon^{\delta-1/2}(tP_1(t)-2/\epsilon)&=&0\\
\lim_{\epsilon\to 0}\lim_{t\to\infty}
\epsilon^{\delta-1/2}(tP_1(t)-2tP_2(t))&=&0
\end{eqnarray}
for every $\delta>0$, and the subscript indicates the one or symmetric
two hole problem.  This is our second main result: {\em The
difference between the escape of the one and two hole problems is determined
to leading order in the small hole limit by the Riemann Hypothesis.}  The
generalised Riemann hypothesis is the equivalent statement for $L$
functions, and implies that corrections are of order $\epsilon^{\delta+1/2}$
in the rational two hole case.
For irrational $\theta$ the above analysis breaks down;
presumably the poles on the
critical line become dense, blocking analytic continuation.  However the
leading order (in $\epsilon$) term can be shown to be $\theta$ independent for
$\theta>0$, as for the rational values given above, as follows. 
For the leading order behavior, the sum
over $n$ can be approximated by an integral, with parts of the
summand replaced by ``mean field'' averages $\langle\rangle$.
If $\theta$ is irrational, the fractional parts are uniformly distributed,
so that we compute
\longv{
\begin{eqnarray}\nonumber
&&\langle g\left(\frac{2\pi}{n}-\theta'-\epsilon\right)+
g\left(\theta'-\epsilon\right)\rangle\\\nonumber
&=&\frac{n}{2\pi}\int_0^{2\pi/n}
\left[g\left(\frac{2\pi}{n}-\theta'-\epsilon\right)+
g\left(\theta'-\epsilon\right)\right]d\theta'\\
&=& \frac{n}{3\pi}\left(\frac{2\pi}{n}-\epsilon\right)^3
\end{eqnarray}
}
\shortv{
\begin{equation}
\langle g\left(\frac{2\pi}{n}-\theta'-\epsilon\right)+
g\left(\theta'-\epsilon\right)\rangle=
\frac{n}{3\pi}\left(\frac{2\pi}{n}-\epsilon\right)^3
\end{equation}
}
We also use $\langle \phi(n) \rangle=6n/\pi^2$ and $\langle \mu(n)\rangle=0$,
so that
\begin{equation}\label{e:irrat}
tP(t)\approx\frac{1}{24\pi^2}\int_0^{2\pi/\epsilon}n^2\frac{6n}{\pi^2}
\left(\frac{2\pi}{n}-\epsilon\right)^3dn=\frac{1}{\epsilon} 
\end{equation}
as required.

\begin{figure}
\begin{picture}(200,180)(90,190)
\includegraphics[width=250pt]{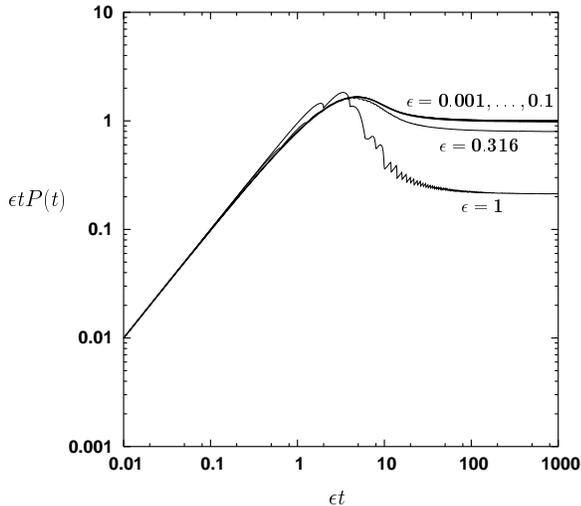}
\end{picture}
\caption{The scaled survival probability $\epsilon tP(t)$
(see Eq.~\protect\ref{e:tP(t)})
appears to approach a limiting function as $\epsilon\to 0$ with $\epsilon t$
held constant.  Here we use $10^8$ random initial conditions, $\theta=1$
(an irrational multiple of $\pi$) and the curves are $\epsilon=10^{-n/2}$
with $n=0\ldots 6$.  At large $\epsilon t$ the function converges
to unity, consistent with Eq.~(\protect\ref{e:irrat}).
\label{f:dist}}
\end{figure}

\begin{figure}
\begin{picture}(200,195)(90,280)
\includegraphics[width=400pt]{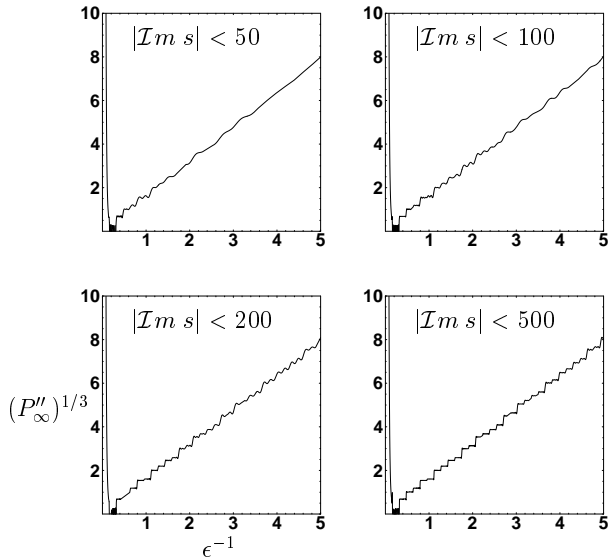}
\end{picture}
\caption{Numerical computation of $P_\infty''(\epsilon)$ for the single hole
case $q=1$, using Eq.~(\protect\ref{e:Ps}) and varying the number of
poles considered on the critical line; real poles are considered for $s>-10$ in
all plots.  From~(\protect\ref{e:tP(t)}) it is seen that the second
derivative of $P_\infty$ is a function with uniformly spaced
steps and constant average gradient in the variables shown on the axes.
Convergence is evident except for very small $\epsilon^{-1}$ for which
more real poles would be required; large
$\epsilon^{-1}$ require more critical poles for the steps to be visible.
\label{f:step}}
\end{figure}

We now present some numerical tests.  The main results of this paper are
exact.  However there are some related questions to
do with rates of convergence of various limits which are of great importance
to numerical or experimental extensions of this work.
The rate of convergence of $tP(t)$ as $t\to\infty$ is considered in
Fig.~\ref{f:dist}.  The convergence of Eq.(\ref{e:Ps}) with the number of
residues is tested in Fig.~\ref{f:step}.  
\longv{
Note that considering
only the real poles would lead to a Taylor series of integer powers, that
is, an analytic function; the complex poles are essential to obtain the
step structure.
}

In conclusion, we computed the escape from a circular billiard with one hole
and related it to the Riemann Hypothesis, computed escape for two holes
(rational case) and related it to the generalized Riemann hypotheses
and obtained the leading order behavior for the irrational two hole case.
Many interesting questions remain, concerning one or two hole escape from
pseudo-integrable, chaotic or mixed billiards, and to the quantum/wave
signature of these systems~\cite{RH02}.  A fuller understanding of
open quantum billiards should have important practical benefits, for
example in the design of microlasers~\cite{Chern03}. 

We would like to thank E. Croot and Z. Rudnick for helpful discussions.
LB was partially supported by NSF grant DMS \#0140165.
CD is grateful for the hospitality of the Center for Nonlinear Science,
Georgia Tech.

\end{document}